\documentclass[aoas,MSNbibl,nameyear,seceqn,dvips]{arximspdf}
\usepackage{graphicx}

\doi{10.1214/12-AOAS568} 
\volume{6}
\issue{4}
\pubyear{2012}
\firstpage{1775}
\lastpage{1794}

\makeatletter
\def\mid{\vert}
\def\O{\mathbf{O}}
\def\T{\mathbf{T}}
\def\p{p_\theta}

\newcommand{\eqref}[1]{(\ref{#1})}
\makeatother

\begin{document}
\begin{frontmatter}

\title{A likelihood-based scoring method for peptide identification
using mass spectrometry}
\runtitle{A likelihood-based score for peptide identification}

\begin{aug}
\author[A]{\fnms{Qunhua} \snm{Li}\corref{}\ead[label=e1]{qunhua.li@psu.edu}},
\author[B]{\fnms{Jimmy K.} \snm{Eng}\ead[label=e2]{engj@uw.edu}}
\and
\author[C]{\fnms{Matthew} \snm{Stephens}\ead[label=e3]{mstephens@uchicago.edu}}
\runauthor{Q. Li, J. K. Eng and M. Stephens}

\affiliation{Penn State University, University of Washington and
University~of~Chicago}

\address[A]{Q. Li\\
Department of Statistics\\
Penn State University\\
326 Thomas Building\\
University Park, Pennsylvania 16802\\
USA\\
\printead{e1}}

\address[B]{J. K. Eng\\
Department of Genome Sciences\\
University of Washington\\
Box 355065\\
Seattle, Washington 98195\\
USA\\
\printead{e2}}

\address[C]{M. Stephens\\
Department of Statistics and Human Genetics\\
University of Chicago\\
Eckhart Hall Room 126\\
5734 S. University Avenue\\
Chicago, Illinois 60637\\
USA\\
\printead{e3}}

\end{aug}

\received{\smonth{11} \syear{2011}}
\revised{\smonth{5} \syear{2012}}

%
\begin{abstract}
Mass spectrometry provides a high-throughput
approach to identify proteins in biological samples. A key step in the analysis
of mass spectrometry data is to identify the peptide sequence that,
most probably, gave rise to each observed spectrum. This is often
tackled using a database search:
each observed spectrum is compared against a large number of
theoretical ``expected'' spectra predicted from candidate peptide
sequences in a database, and the best match is identified using some
heuristic scoring criterion.
Here we provide a more principled, likelihood-based, scoring criterion
for this problem. Specifically, we introduce a
probabilistic model that allows one to assess, for each theoretical
spectrum, the probability that it would produce the observed spectrum.
This probabilistic model takes
account of peak locations \textit{and} intensities, in both observed and
theoretical spectra, which enables incorporation of detailed knowledge
of chemical plausibility in peptide identification.
Besides placing peptide scoring on a sounder theoretical footing,
the likelihood-based score also has important practical benefits: it provides
natural measures for assessing the uncertainty of
each identification, and in comparisons on benchmark data it produced
more accurate peptide identifications than other methods, including SEQUEST.
Although we focus here on peptide identification, our scoring rule
could easily be integrated into any downstream analyses that require
peptide-spectrum match scores.
\end{abstract}

%

\end{frontmatter}

\section{Introduction}\label{S:introduction}

Tandem mass spectrometry (MS/MS) provides a high-throughput approach to
identify proteins in biological samples.
In a typical MS/MS experiment, proteins in the sample are first broken
into short sequences, called peptides, and the resulting mixture of
peptides is subjected to mass spectrometry,
which fragments peptides and generates tandem mass spectra that
contain fragmentation peaks characteristic of their generating peptides
[\citet{coon:05}, \citet{kinter:00}].
A variety of computational methods are then used to process the mass
spectra, with, typically, the ultimate goal being to identify which
proteins and/or peptides are present in the mixture, and to provide
some measure of confidence in these identifications.

The computational pipelines used for processing
these kinds of data can vary considerably, even for analyses that share
the same ultimate goal.
However, one element that plays an important role in the vast majority
of these pipelines is
the need to ``score'' how well each observed spectrum matches a number
of candidate generating peptides. Despite the fact that such
scoring procedures play a key role in all kinds of downstream analyses,
existing scoring procedures are generally fairly simple and ad hoc. In
this paper we develop a more statistically
rigorous, likelihood-based, approach to peptide-spectrum scoring, which,
in the examples considered later, performs better than existing scoring
rules (e.g.,~the Xcorr score in SEQUEST) in discriminating between the
true generating peptide and other candidate peptides.
This scoring rule
could be integrated easily into any downstream
analyses that require peptide-spectrum
match scores, including decoy database search strategies [\citet{elias:07}] and formal statistical modeling approaches for protein
identification [\citet{gerster:10}, \citet{li:10}, \citet{nesvizhskii:03}, \citet{shen:08}].



In brief, our scoring approach, in common with many existing
approaches, has two steps: first,
for a given candidate peptide sequence, we generate a theoretical
``expected'' spectra; second, the observed spectrum is compared with
this theoretical spectrum.
Most existing algorithms [reviewed in \citet{hernandez:06}, \citet{sadygov:04}]
use simple approaches in both these steps. Specifically, they typically
use coarse theoretical spectra containing only predicted locations (not
intensities) of spectral peaks derived from a few major chemical
fragmentation pathways, and score similarity primarily by the matching
of peak locations (again ignoring peak intensities) using ad hoc rules.
The resulting peptide identification procedures are generally rather
inaccurate (typically 70--90\% of the top-scoring peptide
identifications are incorrect [\citet{keller:02}, \citet{nesvizhskii:04}]).

In comparison, our approach attempts to be more sophisticated in both
of these steps. For the first step we make use of the improved
theoretical prediction algorithm from \citet{zhang:04} [see also \citet
{klammer:08}], which incorporates gas phase chemistry mechanisms of
peptides into a kinetic model for peptide fragmentation, to generate
detailed theoretical predicted spectra for any given peptide. These
detailed theoretical spectra contain predicted locations and
intensities of peaks from both major and minor fragmentation pathways,
all of which may help improve accuracy of peptide identification.
Indeed, such\vadjust{\goodbreak} features are commonly used in manual annotation to
validate chemical plausibility of putative peptide identifications
[\citet{sun:07}]. For the second step we develop a novel
likelihood-based scoring rule for this problem. This likelihood-based
approach is based on a probabilistic model for differences between the
theoretical and observed spectrum, in both peak intensities and
locations, and allows for the fact that predicted low-intensity peaks
from minor pathways are more often absent from observed spectra than
are predicted high-intensity peaks from major pathways. It is in this second
step that our work differs from all existing scoring
algorithms, including the few that do make use of complex predicted
spectra [\citet{yen:11}, \citet{zhang:04}], which by comparison use simple, ad
hoc, measures of similarity to compare observed and theoretical spectra.

As we demonstrate on examples later, likelihood-based scoring has some
important practical benefits: it provides
natural measures for assessing the uncertainty of
each identification, and, on benchmark data we consider here,
ultimately improves the accuracy of identification. In addition, it has
the attraction of putting peptide scoring on a sounder theoretical
statistical footing.

The structure of the paper is as follows. We first describe the
generation of theoretical spectra (Section~\ref{SS:prediction}) and the
procedure we use to preprocess both observed and theoretical spectra
(Section~\ref{SS:preprocessing}). Section~\ref{SS:model} describes our
probabilistic model and the methods we use to estimate parameters in
this model and score peptide sequences. In Section~\ref{S:simulation}
we check the effectiveness of these methods on simulated data. In
Section~\ref{2S:ISB} we illustrate the methods using a publicly
available benchmark data set. Section~\ref{2S:discussion} concludes and
discusses future work.

\section{Methods and models}\label{S:method}

\subsection{Refined theoretical spectra and its use in peptide
identification}\label{SS:prediction}

We use the chemical model from \citet{zhang:04} to predict the
theoretical spectra for peptide sequences. This model generates refined
theoretical spectra containing both locations and intensities of the
peaks from comprehensive pathways. For convenience of producing our own
pipeline, we coded this prediction algorithm in Java.
Our implementation produced similar results to Zhang's software for the
examples shown in \citet{zhang:04}, although there were some
quantitative differences (peak heights did not always agree). In as
much as these differences could reflect deficiencies of our
implementation, we note that correcting these deficiencies would be
expected to yield further improvements in performance compared with
those reported below. Our implementation is available on request from
the first author.

\begin{figure}

\includegraphics{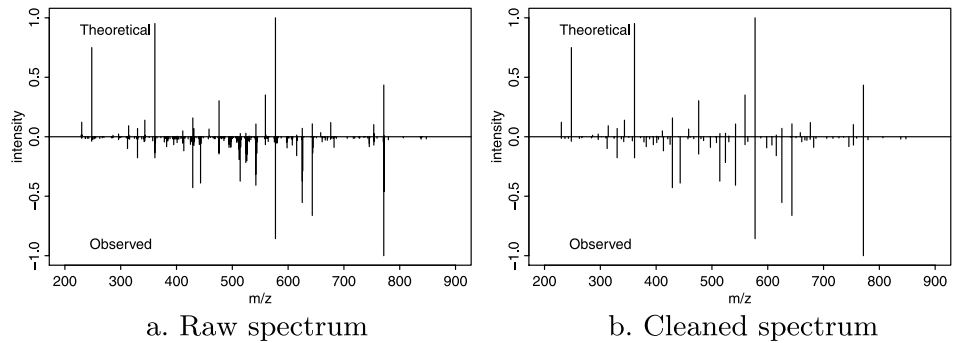}

\caption{Observed and theoretical spectra of a charge 1$+$ peptide
sequence LVTDLTK. In each plot, top panel is the theoretical spectrum
predicted using our implementation based on \protect\citet{zhang:04}, and
bottom panel is the observed spectrum. Each spectrum is rescaled by
dividing by its highest peak for better visualization. \textup{(a)} Raw spectrum.
\textup{(b)} Cleaned spectrum from our preprocessing procedure.}\label{2F:mirror}
\end{figure}

Though there is still marked deviation between theoretical prediction
and observed spectra [e.g., Figure~\ref{2F:mirror}(a)], the more refined
predicted spectra from this model increase the detail with which one
can assess similarity of observed and theoretical spectra.

\subsection{Preprocessing}\label{SS:preprocessing}

Observed tandem mass spectra (specifically, those produced by LCQ or
LTQ instruments) usually contain a large number of clustered peaks and
low-intensity peaks, and have highly variable peak intensities [Figure~\ref{2F:mirror}(a)]. These factors pose challenges for developing
statistical models. For example, clustered peaks often represent
variants from the same fragmentation product (e.g.,~isotopic peaks) and
so are not independent. Preprocessing has been reported to be important
for the accuracy of peptide identification [\citet{sun:07}].

Here we use a novel preprocessing procedure that attempts to distill
the spectra down to the primary signals, normalizes the peak
intensities on all theoretical and observed spectra to a comparable
scale, and stabilizes peak intensities.
In brief, the procedure distills the spectra down to the primary
signals by clustering neighboring peaks and pooling near-by peaks into
a single representative peak (Figure~\ref{2F:mirror}). Peak intensities
in each cleaned spectrum are then normalized by dividing by the 90th
percentile of the intensities of the peaks on the spectrum, to put the
peaks from different spectra on a comparable scale. Finally, the
normalized intensities are transformed by raising to 1$/$4 power to
stabilize the highly variable intensities.
We apply the same procedure to both theoretical and observed spectra
before scoring. The preprocessing steps are described in detail in
Table S1 in the supplementary materials [\citet{li:12}].

\subsection{A probabilistic model}\label{SS:model}

We now outline the probabilistic model that is the central contribution
of this paper. Let $\mathbf{T}=(T_1,\ldots,T_n)$ be a predicted
theoretical spectrum with $n$ spectral peaks, where $T_i=(X_i^t,
Y_i^t)$ denotes the location ($X^t_i$) and intensity ($Y^t_i$) for the
$i$th peak, and $X^t_{i_1} \neq X^t_{i_2}$ if $i_1 \neq i_2$.
Similarly, let $\mathbf{O}=(O_1,\ldots,O_m)$ be an observed spectrum
with $m$ spectral peaks, where $O_j= (X^o_j, Y^o_j)$ and $X^o_{j_1}
\neq X^o_{j_2}$ if $j_1 \neq j_2$. We find it convenient to assume that
the peaks in $\O$ and $\T$ are arbitrarily ordered (i.e., they are
randomly labeled $1,\ldots,n$ in $\T$ and $1,\ldots,m$ in $\O$), rather
than being ordered by location, for example.

If $\mathbf{T}$ and $\mathbf{O}$ are generated from the same peptide
sequence, then we view $\mathbf{O}$ as a distorted (i.e.,~noisy)
realization of $\mathbf{T}$. Our aim here is to define a probability
model $\p(\O| \T)$, depending on a set of parameters, $\theta$,
described in detail below, that captures this distortion. [Actually, in
specifying the probability model we condition on the number of peaks,
$m$, in the observed spectrum, so $\p(\O| \T)$ should read $\p(\O| \T
,m)$, but for notational simplicity we omit the explicit conditioning
on $m$.]

\begin{figure}

\includegraphics{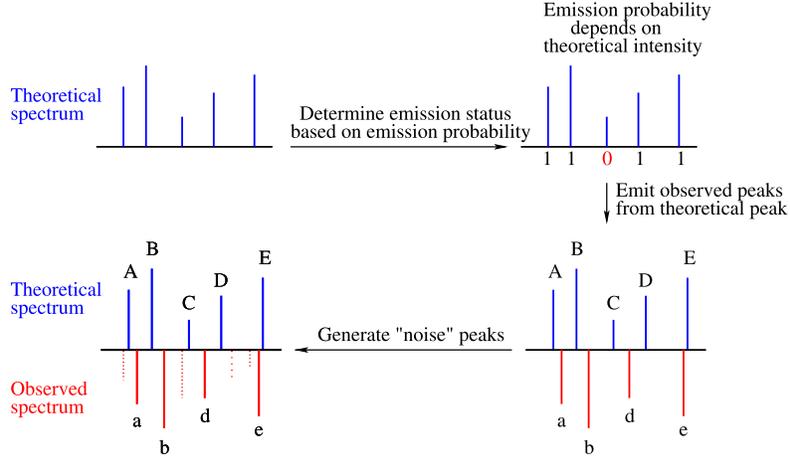}

\caption{The probabilistic model for generating a random observed
spectrum from a given theoretical spectrum, described in
Section \protect\ref{SS:model}.}\label{2F:model}
\end{figure}

To specify this model, consider generating a random ``observed''
spectrum $\O$ from $\p(\O| \T)$ as follows (Figure~\ref{2F:model}):
\begin{longlist}[(1)]
\item[(1)] Each theoretical peak either does or does not ``emit'' an
observed peak, independently for the $n$ theoretical peaks. We use
$p_i$ to denote the probability that the $i$th theoretical peak, which
with slight abuse of notation we abbreviate as $T_i$, emits a peak. We
allow $p_i$ to depend on the intensity of $T_i$, so $p_i =
g(Y^t_i,\theta)$ for some function $g$, defined below. If $T_i$ emits a
peak, then the location of the emitted peak is randomly sampled from a
truncated normal distribution centered at $X_i^t$, and its intensity is
randomly sampled from some distribution $f_1(\cdot; \theta)$.
\item[(2)] Assume that the previous step produces $k$ emitted peaks, where
$k \leq n \leq m$. (Typically, including every case we needed to
consider in practical applications presented here, $n \leq m$.) Now
generate $m-k$ additional ``noise'' peaks, so that the total number of
peaks is $m$. These noise peaks have locations independently randomly
sampled from a uniform distribution across the whole observable $m/z$
range, and intensities independently sampled from a distribution
$f_0(\cdot;\theta)$. Note that, despite their name, these noise peaks
may represent either measurement noise, or genuine peaks that were
simply not included in $\T$ due to limitations of the theoretical
prediction model.
\item[(3)] Randomly label the observed peaks $1,\ldots,m$, uniformly on all
possible labelings.
\end{longlist}

The above process is flexible enough to capture several important
properties of real data. For example, by letting $p_i$ depend on the
intensity of~$T_i$, it captures the fact that high-intensity
theoretical peaks are more likely to have matching observed peaks. And
by allowing $f_1$ to be stochastically larger than $f_0$, it can take
account of the fact that observed peaks that match a theoretical peak
will tend to have higher intensities than other observed peaks (Figure
\ref{2F:mirror}). Here we assume that $g$ is a logistic function, and
use histogram-like density estimates (i.e., piecewise constant
densities) for $f_0$ and $f_1$, with parameters of these functions
being estimated from data
as described below.

The probability $\p(\O| \T)$ captures how probable it is that a
peptide with theoretical spectrum $\T$ would have resulted in the
observed spectrum $\O$, and is thus a suitable scoring function for
comparing different candidate $\T$s to identify the peptide that
created $\O$. However, although $\p(\O| \T)$, described above, is very
easy to simulate from, it is tricky to compute for any given~$\T$,
because we do not observe which theoretical peak, if any, ``emitted''
each observed peak. So computing $\p(\O| \T)$ involves a
computationally-intensive sum over all possibilities.

To formalize this model, let $\mathbf{e}$ denote the unobserved
``emission configuration'' which identifies which theoretical peaks
emitted which observed peaks. Each emission configuration $\mathbf{e}$
determines an emission function $e^t$, with $e^t(i)=j$, ($j=1, \ldots,
m$), if $T_i$ emits $O_j$, and $e^t(i)=0$ if $T_i$ does not emit any
observed peak. Similarly, $\mathbf{e}$ also determines another emission
function $e^o$, with $e^o(j)=i$, ($i=1, \ldots, n$), if $O_j$ is
emitted from $T_i$, and $e^o(j)=0$ if $O_j$ is a noise peak. Note that
$e^t(\cdot)$ and $e^o(\cdot)$ contain the same information.\looseness=1

Now, we can write $\p(\O| \T)$ as a sum over all possible values of
$\mathbf{e}$:
%
\begin{equation}
\label{E:likelihood} \p(\mathbf{O} \mid\mathbf{T} )= \sum
_{\mathbf{e}}\bigl[\p(\mathbf{O} \mid \mathbf{T}, \mathbf{e})\p(
\mathbf{e} |\T)\bigr].
\end{equation}
Here
%
\begin{equation}
\label{E:emission} \p(\mathbf{e} | \T)= \frac{(m-k)!}{m!}\prod
_{\{i:e^t(i) > 0\}} g\bigl(Y^t_i;\theta\bigr) \prod
_{\{i:e^t(i)=0\}} \bigl(1-g\bigl(Y^t_i;
\theta\bigr)\bigr),
\end{equation}
where $k \equiv|\{i\dvtx  e^t(i)>0\}|=|\{j\dvtx  e^o(j)>0\}|$ is the number of
emission peaks; and
%
\begin{eqnarray}
\label{E:ogivent}
\p(\mathbf{O} \mid\mathbf{T}, \mathbf{e}) &=& \biggl(
\frac{1}{r}\biggr)^{m-k}\prod_{\{j: e^o(j)=0\}}f_0
\bigl(Y^o_j\bigr)
\nonumber
\\[-8pt]
\\[-8pt]
\nonumber
&&{} \times\prod_{\{j:e^o(j)>0\}} \bigl[N_T
\bigl(X^o_j; X^t_j,
\sigma^2, w\bigr)f_1\bigl(Y^o_j
\bigr)\bigr],
\end{eqnarray}
where $r$ is the length of the $m/z$ range of the uniform distribution
on noise peaks, and $N_T(x; \mu,\sigma^2,w)$ denotes the truncated
normal distribution, with mean $\mu$, variance $\sigma^2$, truncated at
distance $w$ from the mean (here we assume that $w >0$ is a known
constant reflecting the precision of the instrument).

Each term in the sum \eqref{E:likelihood} is easy to compute. However,
the number of terms is sufficiently large to create computational challenges
(even when we take account of the fact that
each emitted observed peak must be within $\pm w$ of the corresponding
theoretical peak, which does
substantially reduce the number of terms, and provides the primary
motivation for using a truncated normal distribution rather than a
nontruncated normal).
To reduce the computation, we replace the likelihood~\eqref
{E:likelihood} with the complete data likelihood under the most
probable configuration, that is,
%
\begin{equation}
\label{E:complete}
\hat{L}(\theta; \mathbf{O}, \mathbf{T}):=\max_{\mathbf{e}}
\bigl[\p(\mathbf {O} \mid\mathbf{T}, \mathbf{e})\p(\mathbf{e}\mid\mathbf{T})
\bigr].
\end{equation}
The procedure for searching the most probable configuration and
estimating parameters is described in detail in Section~\ref{SS:estimation}.
In general, the complete data likelihood \eqref{E:complete} will be a
good approximation to the likelihood \eqref{E:likelihood} only if the
sum in the latter is dominated by its single biggest term, which will
not always be the case. Nonetheless, simulations (Section~\ref{S:simulation})
and empirical evaluation presented below demonstrate that use of \eqref
{E:complete} produces good performance in practice.

\begin{figure}

\includegraphics{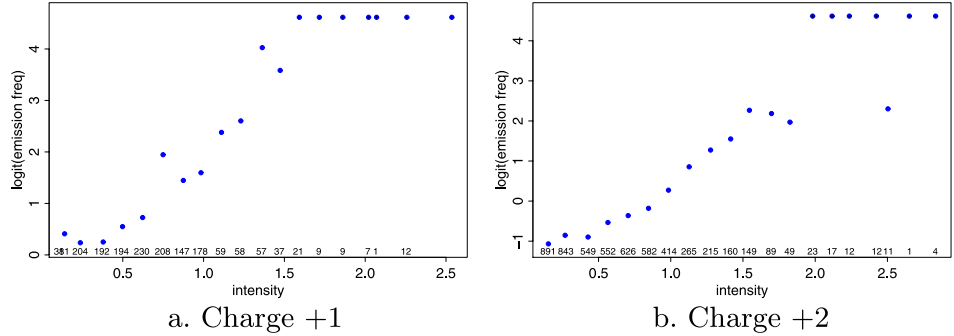}

\caption{Exploratory data analysis shows an approximate linear trend
between theoretical intensities and logit of empirical emission frequencies.
The theoretical intensities are preprocessed as described in Section
\protect\ref{SS:preprocessing}, and are binned (20 bins) with a fixed binwidth.
The emission frequencies are estimated as the proportions of putative
matches (i.e., the observed peaks and the theoretical peaks that locate
less than 2 Daltons apart) in each bin from training data.
To avoid overflow when all peaks in a bin form putative matches, $p$ is
bounded at 0.99 at plotting (i.e., Y-axis is bounded at 4.69). The
number of observations in each bin is marked at the bottom of the plot.}\label{2F:linearity}
\end{figure}

\begin{figure}[b]

\includegraphics{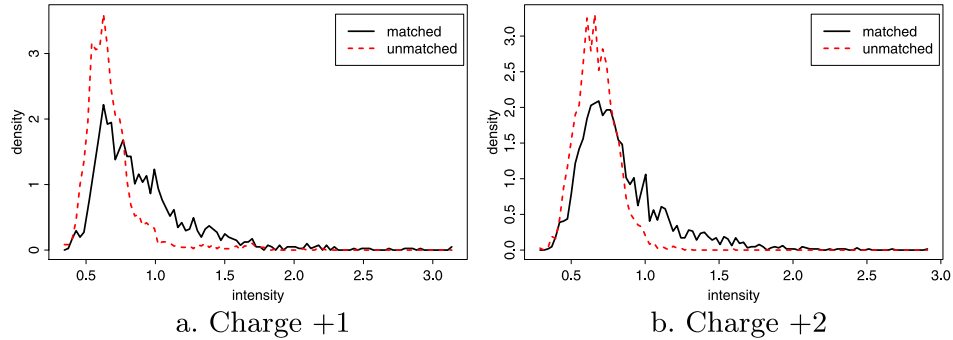}

\caption{Empirical distributions of intensities for peaks on observed
spectra. The emission status of each observed peak is approximated by
whether there exists a theoretical peak that is less than 2~Da apart
from the observed peak: Matched: observed peaks within 2~Da of a
theoretical peak; Unmatched: other observed peaks. The observed
intensities are normalized and transformed as described in
Section \protect\ref{SS:preprocessing}
and are binned (100 bins) with even binwidth for plotting.}\label{2F:observed}
\end{figure}

\subsection{Choice of $g$, $f_0$ and $f_1$}\label{SS:theoretical}

As noted above, we assume that $g$ is a logistic function.
That is, we allow $p_i$ to depend on $y^t_i$ using a logistic regression:
%
\begin{equation}
\label{E:logistic} \log{\frac{p_i}{1-p_i}} = \mu+ \beta y^t_i.
\end{equation}
Here the intercept $\mu$ is assumed to be spectrum-specific to take
account of the variation across spectra, and the slope $\beta$ is
assumed to be common to all spectra.
Exploratory data analysis (e.g., Figure~\ref{2F:linearity})
suggests that a logistic form for $g$ is reasonable for our data.

Empirical evaluations on the intensities of peaks on observed spectra
show that both $f_0$ and $f_1$ are heavy-tailed distributions (Figure
\ref{2F:observed})
where $f_1$ has a heavier right tail than $f_0$. Rather than make
specific parametric assumptions regarding $f_0$ and $f_1$, we allow
them to take flexible shapes by using piecewise-constant densities. For
the data here, we made a specific choice of a piecewise-constant
function of 10 bins, where the highest $1\%$ of the intensities of
observed peaks in the training set is contained in the $10${th} bin
and the remaining $99\%$ are equally distributed in the remaining 9
bins. The same bin boundaries are used for $f_0$ and $f_1$. These
parameters are assumed to be common to all spectra.\eject

%


\subsection{Parameter estimation, scoring and initialization}\label
{SS:estimation}

For the applications presented here we used a supervised approach to
estimate the parameters of our probabilistic model, based on the availability
of a training
set of $N$ spectrum pairs $(\mathbf{O}_s, \mathbf{T}_s)\ (s=1, \ldots, N)$,
where $\mathbf{O}_s$ and $\mathbf{T}_s$ are known to be generated from
the same peptide. However, these parameters could also be estimated in
other ways when training data are not available; see \hyperref[2S:discussion]{Discussion}.

We write $\theta=(\theta_0, \mu_1, \ldots, \mu_N)$ for the parameters
to be estimated, where $\theta_0 =(\sigma^2, \beta, f_0, f_1)$ denotes
parameters shared across all spectra pairs in \eqref{E:complete}, and
$(\mu_1, \ldots, \mu_N)$ are spectrum-specific intercepts defined in
\eqref{E:logistic}. Using the training data, we estimate $\theta=(\theta_0, \mu_1, \ldots, \mu_N)$ by
%
\begin{equation}
\label{theta} \hat{\theta}=\operatorname{argmax} \prod
_s \hat{L}(\theta; \mathbf{O}_s,
\mathbf{T}_s).
\end{equation}

When scoring a spectrum $\mathbf{T}$ for an observed spectrum $\mathbf
{O}$, we then compute the score function
%
\begin{equation}
\label{score} S(\mathbf{T}; \mathbf{O}):=\max_{\mu} \hat{L}(\hat{
\theta}_0, \mu; \mathbf{O}, \mathbf{T}),
\end{equation}
using $\hat{\theta}_0$ estimated from the training stage. The spectra
at different charge states are trained and scored separately.

Because the term $\hat{L}$ defined in \eqref{theta} and \eqref{score}
involves both $\mathbf{e}$ and $\theta$, the maximization involves
simultaneously maximizing the configuration $\mathbf{e}$ and
parameters. Detailed steps are described in Table~\ref{2T:training} in
the \hyperref[app]{Appendix}.
As a configuration is updated only when the likelihood is increasing,
this procedure guarantees the likelihood will be nondecreasing
throughout the procedures. This procedure usually converged within 30
iterations for the data we tested.

\subsection{Uncertainty of identifications}\label{SS:uncertainty}

One advantage of our likelihood-based scoring approach is that it leads
naturally to an assessment of the confidence that a given peptide
generated a given spectrum. Specifically, if we assume that
exactly one of the candidates generated the observed spectrum, and all
are equally likely a priori, then by Bayes' theorem
the probability that $\mathbf{T}_t$ generated $\mathbf{O}$ is given by
%
\begin{equation}
\label{posterior} P(\mathbf{T}_t \mid\mathbf{O}) = \frac{p(\mathbf{O}\mid\mathbf
{T}_t)}{\sum_{i\in C_o}p(\mathbf{O}\mid\mathbf{T}_i)},
\end{equation}
where $C_o$ is the collection of candidate sequences for the observed
spectrum~$\mathbf{O}$.
In practice, we use the scores
$S(\mathbf{T}; \mathbf{O})$ in place of $p(\mathbf{O}\mid\mathbf{T})$
to approximate this expression.

A nice feature of \eqref{posterior} is that it weighs the evidence for
different candidates directly against each other, rather than against
some null hypothesis,
which often requires\vadjust{\goodbreak} a large number of candidates to obtain reasonable
estimates of uncertainty, as in, for example, \citet{fenyo:03}, \citet{klammer:09}.
Thus, for example, if several good candidates provide similarly good
matches to the observed spectrum $\mathbf{O}$, then we could not be
confident which candidate generates $\mathbf{O}$ and this uncertainty
is appropriately reflected in \eqref{posterior}. [On the other hand,
because equation \eqref{posterior}
is based on an assumption that exactly one of the candidates produced
the observed spectrum,
it does not incorporate uncertainty due, for example, to the
possibility that the database search
entirely missed the correct spectrum. Empirically, however, we have
found that the absence of the real sequence from the list of candidates
tends not to cause a problem: in such cases confidence measures of all
candidates tend to be low.]

\section{Simulation}\label{S:simulation}

Here we use simulation studies to examine the performance of our
approach, particularly to assess the accuracy of estimating parameters
using the complete data likelihood (\ref{E:complete}) rather than the
actual likelihood (\ref{E:likelihood}). To do so, we generate observed
spectra from theoretical spectra using the probabilistic model
described in Section~\ref{SS:model},
then estimate parameters and emission statuses by maximizing the
complete data likelihood (\ref{E:complete})
using the estimation procedure in Section~\ref{SS:estimation}.

In an attempt to generate realistic simulations, we first estimate
parameters from a training data of 50 charge $+$1 and 50 charge $+$2
spectra (described in Section~\ref{2S:ISB}) for each charge state, using the estimation procedure described in
Section~\ref{SS:estimation}.
We then simulate one observed spectrum from each theoretical spectrum
in the training set using the estimated parameters, with $0.9n$ noise
peaks on each observed spectrum, where $n$ is the number of theoretical
peaks. We estimate the parameters and evaluate the accuracy of
estimated emission status for each peak on the observed and theoretical
spectra. The simulation is repeated 100 times.

The estimated parameters (Table~\ref{2T:simu-result}) are close to the
true values, which indicates the complete data likelihood at the most
probable emission configuration provides adequate parameter estimates.

\begin{table}
\caption{Parameter estimation and accuracy of estimated emission
status in simulated data. $\operatorname{CE}_T$~is the proportion of
misclassified emission labels for peaks on the theoretical spectra
after estimation, $\operatorname{CE}_O$ is the proportion of misclassified
emission labels for peaks on the observed spectra after estimation.
Each simulation consists of 50 theoretical spectra and their
corresponding observed spectra simulated from the probabilistic model.
Mean and standard deviation are computed based on 100 simulations}\label{2T:simu-result}
\begin{tabular*}{\textwidth}{@{\extracolsep{\fill}}lcccc@{}}
\hline
& \multicolumn{2}{c}{\textbf{Charge} $\bolds{+1}$} & \multicolumn{2}{c@{}}{\textbf{Charge} $\bolds{+2}$}
\\[-4pt]
& \multicolumn{2}{c}{\hrulefill} & \multicolumn{2}{c@{}}{\hrulefill} \\
& \textbf{True} & \textbf{Estimated} & \textbf{True} & \multicolumn{1}{c@{}}{\textbf{Estimated}} \\
& \textbf{parameter} & \textbf{from} $\bolds{\hat{L}}$ & \textbf{parameter} & \multicolumn{1}{c@{}}{\textbf{from} $\bolds{\hat{L}}$}\\
\hline
$\mu$ & $-1.240$ &$-1.052$ (0.119) & $-5.060$ & $-5.037$ (0.145)\\
$\beta$ & \phantom{$-$}2.970 & \phantom{$-$}2.929 (0.188) & \phantom{$-$}4.740 & \phantom{$-$}4.785 (0.150)\\
$\sigma$ & \phantom{$-$}0.390 & \phantom{$-$}0.382 (0.008) & \phantom{$-$}0.160 & \phantom{$-$}0.157 (0.003)\\
$\operatorname{CE}_{T}$ & -- & \phantom{$-$}0.046 (0.004) & -- & \phantom{$-$}0.014 (0.002)\\
$\operatorname{CE}_{O}$ & -- & \phantom{$-$}0.049 (0.004) & -- & \phantom{$-$}0.028 (0.003)\\
\hline
\end{tabular*}
\end{table}

\section{Applications}\label{2S:ISB}

\subsection{ISB data}

To illustrate our approach, we applied it to a widely-used standard
protein mixture, known as the ISB data, for assessing peptide
identification [\citet{keller:02b}].
This data set consists of the MS/MS spectra generated from a sample
composed of trypsin digest of 18 purified proteins, including 504
charge $+$1 spectra, 18,496 charge $+$2 spectra and 18,044 charge $+$3 spectra.
The spectra in the data set have been analyzed using SEQUEST [\citet
{eng:94}], a commonly-used software for peptide identification, and a
list of 10--11 top-ranked candidates selected by SEQUEST was provided
for each spectrum. For a subset of spectra,\vadjust{\goodbreak} hand-curation (i.e.,~manual
inspection) has confirmed that the top-ranked peptide assignment from
SEQUEST is correct.
This subset, which we refer to as the ``hand-curated dataset'' in what
follows, consists of 125 $+$1 spectra, 1640 $+$2 spectra and 1010 $+$3
spectra. The experimental procedures are described in \citet{keller:02b}.

Because we implemented Zhang's prediction model only for charge $+$1 and
$+$2 peptides, here we consider only the observed spectra at these charge
states, though our scoring method could also be applied to spectra at
other charge states [\citet{zhang:05}]. As the resolution of the
instrument to generate the spectra in this data set is about 2~Da [\citet
{wan:06}], we set $w=2$~Da [e.g., in (\ref{E:ogivent}) and Table~S1 in
the supplementary materials [\citet{li:12}].

Because of the computational cost involved in predicting theoretical
spectra, the comparison is carried out on only the top 10 candidates
selected by SEQUEST rather than all the candidates in the entire
database. That is, we effectively assess the accuracy of a two-stage
procedure that first selects candidates using SEQUEST, and then refines
the ranking of the candidates shortlisted by SEQUEST using our
likelihood-based score and the similarity index, respectively. Both
theoretical spectra and observed spectra are preprocessed using the
procedure described earlier (Table S1 in the supplementary materials
[\citet{li:12}])
prior to scoring with our method and the similarity index.

\subsection{Other methods compared}

We compare the results from our scoring method with the Xcorr score
from SEQUEST and a similarity index (I) in \citet{zhang:04}. SEQUEST is
one of the most widely-used software
for peptide identification. It scores candidate peptides using coarse
theoretical spectra and reports multiple scores\vadjust{\goodbreak} for each candidate peptide.
XCorr score is the main filter score from SEQUEST, defined as $\operatorname{Xcorr}=
R_0-\sum_{i=-75}^{i=75}R_i/151$, where $R_i$ is the cross-correlation between
the theoretical spectrum and an observed spectrum with lag $i$ [\citet{eng:94}].
The similarity index is defined as
$I = \frac{\sum_{i} \sqrt{y^o_i y^t_{i}}}{\sqrt{\sum_{i}y_i^o}\sqrt{\sum_{i}y_i^t}}$, where $y^o_i$ and $y^t_i$ are the peak intensities at
(discretized) $m/z$ location $i$ on observed and theoretical spectra,
respectively. It
was originally proposed in \citet{zhang:04} for assessing the
similarity between a refined theoretical spectrum generated from the
kinetic model in \citet{zhang:04}
and its corresponding observed spectrum, and recently was used, in
conjunction with other heuristic rules, to validate top-ranked
identifications made by common database search algorithms using refined
theoretical predictions [\citet{sun:07}, \citet{yu:10}].\vspace*{-3pt}

\subsection{Evaluation on the curated data set}\label{2SS:curated}

We first evaluate the performance of our method and the similarity
index on the hand-curated data set. Because, by construction, this data
set includes only spectra that were correctly identified by SEQUEST, we
cannot make meaningful comparisons with SEQUEST on these data.
For our method, for each charge state, 50 observed spectra are randomly
selected as training data, and the remaining spectra are used for
testing (resulting in test sets of size 75 for charge $+$1 and 1590 for
charge $+$2).
No training is needed for computing the similarity index.
It is known that mass spectrometry has difficulty distinguishing
several sets of amino acids due to their close or identical mass.
Specifically, Ile and Leu have identical mass, and Lys is difficult to
distinguish from Gln. To allow for these undistinguishable variants, in
assessing each method's performance on this subset, we call an
identification correct if the peptide candidate receiving the highest
score agrees either with the hand-curated choice or with an
undistinguishable variant (i.e., a variant that swaps Ile with Leu
and/or Lys with Gln).

\begin{table}[b]
\tabcolsep=0pt
\caption{Correct identification rate on the curated ISB data set. The
confident subset consists of testing spectra whose top candidates are
highly confident, that is, $P(\mathbf{T}_{\mathrm{top}}\mid\mathbf{O}) \geq99\%$}\label{2T:ISB}
\begin{tabular*}{\textwidth}{@{\extracolsep{\fill}}lcccc@{}}
\hline
& & \textbf{Charge} $\bolds{+1}$ & \textbf{Charge} $\bolds{+2}$ & \multicolumn{1}{c@{}}{\textbf{All}} \\
\hline
Likelihood score $(S)$ & train & 94.0\% ($n=50$) & 100.0\% ($n=50$)\phantom{000} &
97.0\% ($n=100$)\phantom{0}\\
& test & 93.3\% ($n=75$) & 96.8\% ($n=1590$) & 96.6\% ($n=1665$)\\
& confident subset & 98.3\% ($n=58$) & 98.7\% ($n=1492$) & 98.7\%
($n=1550$)\\[3pt]
Similarity index $(I)$ & test & 78.7\% ($n=75$) & 88.0\% ($n=1590$) &
87.6\% ($n=1665$)\\
\hline
\end{tabular*}
\end{table}

\textit{Average identification accuracy}.
Table~\ref{2T:ISB} summarizes identification accuracy for these data.
Our model correctly identifies most spectra (96.6\% of the spectra in
the test set)
and performs markedly better than the similarity index (87.6\% correct
on test set).\vadjust{\goodbreak}

\textit{High-confidence subsets and calibration of posterior
probabilities}.
Besides providing accurate average performance,
two other features of a method are desirable. First, it should provide
a meaningful ranking of confidence in different identifications. In
particular, it would be helpful
if the method were able to identify a subset of high confidence
identifications that are highly likely to be correct. Second, it should
provide a calibrated assessment of confidence in each individual
identification: in our case, one would like the probabilities assigned to
individual identifications to be calibrated, so that, for example, of
identifications assigned 50\% probability of being correct, around half
are actually
correct.

\begin{figure}

\includegraphics{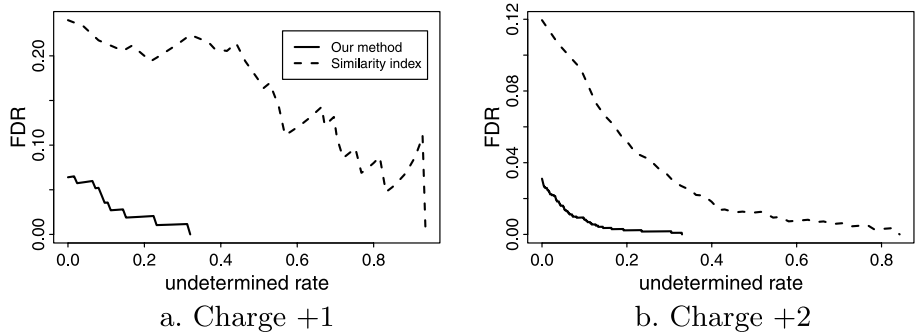}

\caption{False discovery rate versus undetermined rate for the test
data in ISB curated data set. Shown is the comparison between posterior
probability from our method (solid line) and the difference in the
similarity index between the best and second-best identifications
(dashed line).}\label{2F:ISB-uncertainty}\vspace*{3pt}
\end{figure}

For these data, our method exhibits both of these desirable properties.
First, among identifications
scored with the highest confidence by our method ($P(\mathbf{T} \mid
\mathbf{O})\geq0.99$), 98.7\% of spectra are correctly identified
(Table~\ref{2T:ISB}). Of course, restricting attention to
this class of high-confidence calls reduces the overall number of
spectra identified: in this case, 93.1\% of spectra fall into our
high-confidence category, so when we use a calling threshold of $0.99$,
$6.9\%$ of spectra are ``undetermined.''
Figure~\ref{2F:ISB-uncertainty} shows the general trade-off between
identification accuracy (actually, False Discovery Rate) and
undetermined rate, as
the calling threshold changes. For comparison, Figure \ref
{2F:ISB-uncertainty} also shows the same trade-off for the similarity
index, with confidence in each call measured by the difference in the
similarity index between the best and second-best identifications.
Although the similarity index is able to provide some
meaningful ranking of confidence in each call---as indicated by the
lower FDR at more stringent thresholds---the FDR at any given
undetermined rate
is consistently higher than for our method.

\begin{figure}
\includegraphics{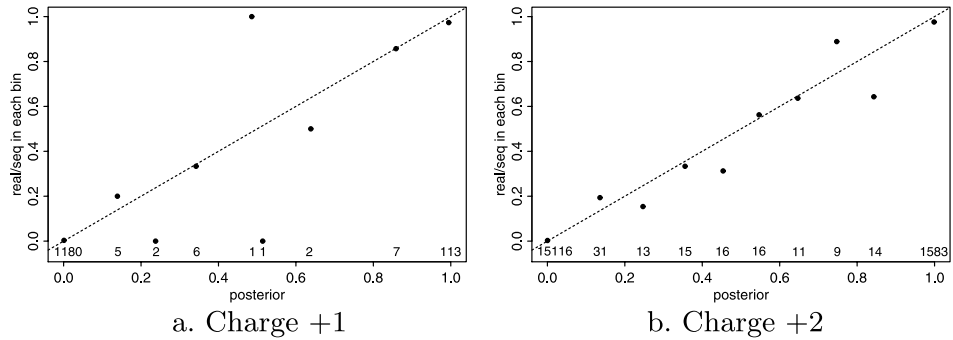}

\caption{Calibration of posterior probabilities in ISB curated data
set. The observations are binned by the assigned probabilities. For
each bin, the assigned probabilities (X-axis) are compared with the
proportion of identifications that are actually correct (Y-axis). The
dashed line marks perfect calibration. The number of observations at
each point is marked at the bottom of the
plot.}\label{2F:calibrate}\vspace*{-3pt}
\end{figure}

Turning to calibration, Figure~\ref{2F:calibrate} shows the calibration
of posterior probabilities from our model. To produce this plot, we
took all candidate sequences in this data set (not just the top-ranked
sequences) and grouped them into bins by their posterior probabilities.
Within each bin we compared the posterior probabilities with the
empirical correct identification rate. The approximate linear trend in
Figure~\ref{2F:calibrate} shows that, for these data, our method
provides reasonably well calibrated probability assessments in both
charge states.
The ability to produce well-calibrated
probabilities of correct identifications is
a potential
advantage of using likelihood-based scoring rules such
as the one we present here, compared with similiarity-based scoring
rules that do not naturally lead to probabilistic assessments of
correctness.\vspace*{-3pt}

%

\subsection{Comparisons with SEQUEST: The benefit of refining
identifications}\label{2SS:rerank}

In this section we evaluate the benefit of using
our likelihood-based score to refine identifications
made by SEQUEST, by comparing the results of our method with
(unrefined) SEQUEST results, and with the similarity index.
Again, because the hand-curated data consists only of identifications
that are correctly identified by SEQUEST, it cannot
be used to make meaningful comparisons with SEQUEST. Instead, we form a
different subset of the ISB data for comparison,
based on the fact that these spectra were generated from a known
mixture of proteins. Specifically, we
take the subset of the (test-set) spectra whose top 10 candidates
selected by SEQUEST include at least one subsequence of a constituent
protein in the known protein mixture. The resulting subset contains 504
charge $+$1 spectra and 3669 charge $+$2 spectra. When assessing a peptide
identification for each spectrum, the assumption we make,
standard in this context, is that the identified
peptide is correct if and only if its amino acid sequence is a
subsequence of a constituent protein in the known mixture.

\begin{table}
\tabcolsep=0pt
\caption{Correct identification rate for the spectra whose top 10
candidate sequences\break selected by SEQUEST include subsequence(s) of the
constituent proteins\break in the ISB data. The confident subset consists of
testing spectra whose top candidates are~highly confident, that is,
$P(\mathbf{T}_{\mathrm{top}}\mid\mathbf{O}) \geq99\%$}\label{2T:rerank-all}
\begin{tabular*}{\textwidth}{@{\extracolsep{\fill}}lcccc@{}}
\hline
& & \textbf{Charge} $\bolds{+1}$ & \textbf{Charge} $\bolds{+2}$ & \multicolumn{1}{c@{}}{\textbf{All}} \\
\hline
Likelihood score $(S)$ & test & 86.9\% ($n=504$) & 87.7\% ($n=3669$) &
87.6\% ($n=4173$)\\
& confident subset & 99.1\% ($n=346$) & 98.0\% ($n=3008$) &98.1\%
($n=3354$)\\[3pt]
SEQUEST & test & 68.1\% ($n=504$) & 82.0\% ($n=3669$) & 80.2\% ($n=4173$)\\[3pt]
Similarity index $(I)$ & test & 60.7\% ($n=504$) & 76.3\% ($n=366$9)&
74.4\% ($n=4173$) \\
\hline
\end{tabular*}
\end{table}

\begin{figure}[b]

\includegraphics{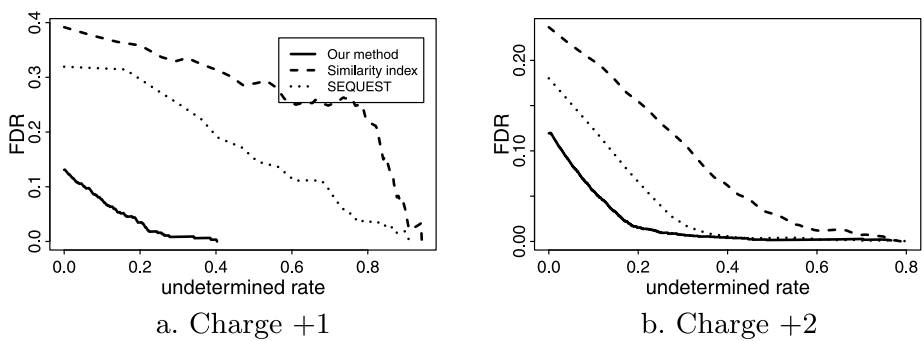}

\caption{False discovery rate versus undetermined rate for the spectra whose top
10 candidates selected by SEQUEST include subsequence(s) of the
constituent proteins in the ISB data. Shown is the comparison between
posterior probability computed using our method (solid line), the
difference in the similarity index between the best and second-best
identifications (dashed line) and Delta Cn of SEQUEST (dotted line),
which is the normalized difference between the top two Xcorr values.
Xcorr performs worse than Delta Cn in this data set, thus it is not shown.%
}\label{2F:top10}
\end{figure}

%

In this data set, our method correctly identifies substantially more
real peptides than SEQUEST and the similarity index (I) for both charge
$+$1 and charge $+$2 spectra\vadjust{\goodbreak} (Table~\ref{2T:rerank-all}). Among them, our
method not only correctly identifies most spectra that SEQUEST
correctly identifies, but also identifies many of the spectra that are
misidentified by SEQUEST. Similar to the hand-curated data set, here
our method
is also able to identify a subset of high-confidence high-accuracy
identifications (e.g., 98.1\% of spectra are correctly identified on
this subset, see Table~\ref{2T:rerank-all}) and provides a
substantially lower false discovery rate than both SEQUEST and the
similarity index at any given undetermined rate (Figure~\ref{2F:top10}).

The gain of using our method to refine SEQUEST
results is noticeably higher for charge $+$1 spectra than for charge $+$2
spectra. \citet{sun:07} also observed a higher gain in charge $+$1
spectra when using the more refined theoretical spectra for peptide
identification using similarity index in conjunction with a series of
heuristic rules. They interpreted this as being due to the fact that
charge $+$1 spectra tend to follow minor fragmentation pathways, which
are excluded in the coarse theoretical spectra but included in the
refined theoretical spectra,
more frequently than charge $+$2 spectra
[\citet{wysocki:00}], so refined theoretical spectra tend to fill in
more information that coarse theoretical spectra miss on charge $+$1 spectra.

We also note that the similarity index identifies fewer real peptides
than SEQUEST in our study. This is different from the results in \citet
{sun:07}, who showed more favorable performance for the similarity
score compared with SEQUEST, when refining the top-ranked peptides
identified by SEQUEST, and MASCOT, another widely-used peptide
identification algorithm, using the similarity scores, in conjunction
with other heuristic rules. This disparity could be attributed to a
number of differences between their study and ours, such as selection
criteria for test spectra (their selection criteria for test spectra
enrich for spectra that are easy to discriminate and also involve
filtering using the similarity index itself), choice of test data sets,
preprocessing procedures (including data transformation) and their use
of additional heuristic rules.

\section{Discussion}\label{2S:discussion}

We have developed a likelihood-based scoring approach to peptide
identification using a database search. This method is based on a
rigorous statistical model, which provides a flexible framework to
model the fine details and noise structure in the spectra. By taking
account of multiple sources of noise in the spectra using a model-based
approach, the method makes use of the information of peak intensities
on both observed spectra and theoretical spectra predicted by
sophisticated chemical principles, in addition to peak locations, in
scoring peptide-spectrum matches. Moreover, the use of a
likelihood-based score leads naturally to an
assessment of the probability that each individual
identification is correct. In the ISB data we examined
here, the probabilities produced by our method were
well-calibrated and produced better identification accuracy than
SEQUEST or the similarity index.

Our results confirm that finer-scale spectra
predicted from comprehensive fragmentation pathways
can provide valuable information for
peptide identification [\citet{sun:07}] and demonstrate the potential to
improve the accuracy of spectra matching by modeling these structures.
Similar improvement in peptide identification was also observed in
\citet{klammer:08}, who developed a probabilistic model of peptide
fragmentation chemistry using the dynamic Bayesian network (DBN) and
identified peptides using the features learned from DBN using the
support vector machine (SVM). Similar to \citet{zhang:04}, \citet
{klammer:08} incorporated peptide fragmentation chemistry using the
widely accepted mobile proton model [\citet{dongre:96}, \citet{wysocki:00}]. It
then trained DBNs using positive and negative spectra and generated a
set of DBNs to capture the probabilistic relationships governing
fragment intensities. Unlike \citet{zhang:04}, it does not produce a\vadjust{\goodbreak}
theoretical spectrum for each peptide candidate; instead, it yields for
each peptide-spectrum match a vector of features from each DBN to be
discriminated by the SVM. One advantage of the likelihood-based scoring
method over the SVM approach is that it can assess the uncertainty of
identification for each identified peptide relative to other candidate
peptides for the same observed spectra.

Although we have focussed here on using the likelihood-based score for
improving accuracy of
peptide identification by peptide database search,
it could also be usefully integrated into many other proteomics
analyses that exploit such scores.
For example, our score could be easily applied to a spectral library
search [\citet{lam:07}], where query spectra are identified by matching
to a library of previously annotated observed spectra. Here the fact
that our score models the fine structure on both query and library
spectra should
be expected to improve accuracy compared with
simpler scoring algorithms. Similarly, peptide-spectrum match scores
from our method
could be used as input to software that use such
scores in downstream analyses, such as identifying the proteins that
are likely present in a mixture [\citet{gerster:10}, \citet{keller:02}, \citet{li:10}, \citet{nesvizhskii:04},
\citet{shen:08}]. The improved performance we observed for
the likelihood-based score in the peptide identification problem,
compared with other scoring rules, should be expected to translate to
improved accuracy of downstream protein identification.

\begin{table}[b]\vspace*{-4pt}
\caption{Procedure for estimating parameters and
scoring}\label{2T:training}\vspace*{-3pt}
\centering
\begin{tabular*}{\textwidth}{@{\extracolsep{\fill}}c@{}}
\hline
\begin{minipage}{\textwidth}
Let $(T, O)$ denote a pair of theoretical and observed spectra in the
training set. For each such pair, do 1(a) and 1(b).\vspace*{-6pt}
\begin{enumerate}[(1)]
\item[(1)] Initialization: %
\begin{enumerate}[(a)]
\item[(a)] Recall, from the main paper, that $e$ denotes the unobserved
``emission configuration'' that maps each peak in $T$ to the peak it
gave rise to in $O$ (or to no peak if it gave rise to no observed
peak). Here we generate a set $E$ containing all possible values for $e$.
\begin{enumerate}[(iii)]
\item[(i)] For each $T_i$, $i=1, \ldots, n$, find $\hat{e}^o(\{i\})$.
\item[(ii)] If $\hat{e}^o(\{i_1\}) \cap\hat{e}^o(\{i_2\}) \neq\varnothing$
for some $i_1$ and $i_2 \in\{1, \ldots, n\}$, merge both the index
sets and the mapped sets, and obtain $\hat{e}^o(\{i_1, i_2\})=\hat
{e}^o(\{i_1\}) \cup\hat{e}^o(\{i_2\})$.
\item[(iii)] Repeat merging, until all mapped sets are mutually exclusive.
Suppose $G$ mutually exclusive sets are obtained with corresponding
sets of theoretical peaks indexed by $I_1, \ldots, I_G$.
The emission configurations of peaks within each putative emission set
$(I_g, \hat{e}^o(I_g))$, $g=1, \ldots, G$, are determined only by peaks
within the set and are independent of peaks in other sets.
\item[(iv)] Let $e_g$ denote the emission configuration $e$ restricted to the
set of theoretical peaks in $I_g$. Thus, $e_g$ maps each theoretical
peak in $I_g$ to $\hat{e}^o(I_g)$. Enumerate all possible values for
$e_g$, and call this set $E_g$.
\end{enumerate}
\item[(b)] Generate an initial configuration to start Step 2:
\begin{enumerate}[(i)]
\item[(i)] For each putative emission set $g$, assign $e_g^0=\operatorname
{argmin}_{i \in I_g, j \in\hat{e}^o(I_g)} |T_i - O_j|$. Denote the
initial configuration as $\mathbf{e}^0=(e^0_1, \ldots, e^0_G)$.
\end{enumerate}
\end{enumerate}
\item[(2)] Maximization:\\
In this section we use the subscript $s$ to label the different spectra
in the training set. So the training set consists of pairs $\{(T_s,
O_s)\dvtx  s=1, \ldots, S\}$ and $e_s$ denotes an emission configuration
mapping peaks in $T_s$ to peaks in $O_s$.

Alternate 2(a) and 2(b) until the log-likelihood $\sum_{s=1}^N \log
L(\theta_0, \mu_1, \ldots, \mu_N)$ converges:
\begin{enumerate}[(a)]
\item[(a)] Estimate spectrum-nonspecific parameters $\theta_0$:
\begin{enumerate}[(i)]
%
\item[(i)]$\hat{\theta}_0^{(t)}= \operatorname{argmax}_{\theta_0} \sum_s \log
L((\theta_0, \mu_1, \ldots, \mu_N) \mid\mathbf{O}_s, \mathbf{T}_s,
\mathbf{e}_s^{(t)})$) for current $\mathbf{e}^{(t)}_s=(e^{(t)}_{s, 1},
\ldots,  e^{(t)}_{s, G})$.
\end{enumerate}
\item[(b)] Update configuration and estimate $\mu_s$: \label{2step:2}\\
For each pair of spectra $\mathbf{T}_s$ and $\mathbf{O}_s$, repeat
2b(i-ii) until $\log L((\hat{\theta}_0, \mu_s)\mid\mathbf{O}_s, \mathbf
{T}_s, \mathbf{e}^{(t)}_s)$ converges.
\begin{enumerate}[(ii)]
\item[(i)] Generate a random permutation $\phi=(\phi_1, \ldots, \phi_G)$ of
$(1, \ldots, G)$.
\item[(ii)] Repeat for $g=1$ to $G$:
\begin{enumerate}[(A)]
\item[(A)] Fix current configurations $e^{(t)}_{s, \phi_1}, \ldots,
e^{(t)}_{s, \phi_{g-1}}, e^{(t)}_{s,\phi_{g+1}}, \ldots, e^{(t)}_{s,\phi_G}$.
For each $e_{s,\phi_g} \in E_{\phi_g}$, define $\mathbf{e}_{s, \phi
_g}^{(t)}=e^{(t)}_{s, \phi_1}, \ldots, e^{(t)}_{s,\phi_{g-1}}, e_{s,\phi
_g}, e^{(t)}_{s,\phi_{g+1}}, \ldots, e^{(t)}_{s,\phi_G}$, compute $\hat
{\mu}_s$= $\operatorname{argmax}_{\mu_s} L((\theta_0, \mu_s)\mid\mathbf{O}_s,
\mathbf{T}_s, \mathbf{e}_{s, g}^{(t)})$.
\item[(B)] Update $e^{(t+1)}_{s, \phi_g}=\operatorname{argmax}_{e_{s, \phi_g} \in
E_g}L((\theta_0, \hat{\mu}_s)\mid\mathbf{O}_s, \mathbf{T}_s, \mathbf
{e}_{s, \phi_g}^{(t)})$
\end{enumerate}
\end{enumerate}
\end{enumerate}
\end{enumerate}
\end{minipage}
\\
\hline
\end{tabular*}
\end{table}

While the work described here provides a solid foundation for a
rigorous statistical approach to the problem of matching spectra
to their generating peptides, there remain many opportunities for
further development and refinement. For example, one issue that would need
to be tackled in practical applications is how to estimate parameters
of the model in the absence of a training set. While simply using
parameters estimated from the ISB data used here might perform
adequately in some cases, one could almost certainly do better
using data generated in the specific context of the experiment to be analyzed.
One simple possibility worth investigating would be to use the most
confident matches identified by a simpler approach, such as SEQUEST, as
a training set;
more statistically rigorous approaches, based, for example, on using an
EM algorithm [\citet{dempster:77}] to learn from unlabeled data, could
also be developed.
While the need
to estimate parameters may initially seem like a drawback, it is also
responsible for an important advantage of likelihood-based scores:
specifically, it makes the likelihood-based score easily adaptable to
the varying characteristics
of spectra under, for example, different charge states and different
machine instrumentation and settings.

Our method requires more computing than some other methods, such as the
similarity score and Xcorr, because it involves optimizing the
likelihood. The computational cost for identifying each peptide
spectrum match is linear in the number of peaks on a cleaned spectrum.
With our prototyping implementation\vadjust{\goodbreak} in R, it takes on average 0.6~s and
1.4~s to evaluate each peptide spectrum match for charge 1$+$ and charge
2$+$ spectra, respectively. An implementation in C is expected to
significantly improve the speed and makes it more suitable for
practical use. A software implementation of the scoring algorithm
described here is available from the first author on request.\vadjust{\goodbreak}

\begin{appendix}
\section*{Appendix: Estimation and scoring procedure}\label{app}

For any subset $I$ of theoretical peaks, let $\hat{e}^o(I)$ denote the
set of observed peaks that could have been generated by the theoretical
peaks in $I$. That is,
%
\begin{equation}
\hat{e}^o(I)= \bigcup_{i \in I} \bigl\{j \in
\{1, \ldots, m\}\dvtx |T_i -O_j| \leq w\bigr\},
\end{equation}
where $I$ is an index set for theoretical peaks. Also, define
%
\begin{equation}
L\bigl((\theta_0, \mu_1, \ldots, \mu_N)
\mid\mathbf{O}, \mathbf{T}, \mathbf {e}\bigr)=p_{\theta}(\mathbf{O}\mid
\mathbf{T}, \mathbf{e})p_{\theta}(\mathbf {e}\mid\mathbf{T}).
\end{equation}

Table~\ref{2T:training} describes the procedure for estimating
parameters and scoring. In the training stage, the entire procedure is
carried out. When scoring spectra in the test set, step 2(a) is omitted
and $\hat{\theta}_0$ estimated from the training stage is used.
\end{appendix}

\section*{Acknowledgments} We would like to thank two anonymous
reviewers and the editor for their constructive comments on improving
the quality of the paper.

\begin{supplement}[id=suppA]\label{suppA}
\stitle{Preprocessing procedure}
\slink[doi]{10.1214/12-AOAS568SUPP} 
\slink[url]{http://lib.stat.cmu.edu/aoas/568/supplement.pdf}
\sdatatype{.pdf}
\sdescription{We describe the preprocessing steps in this supplement.}
\end{supplement}


\printaddresses


\begin{thebibliography}{28}

\bibitem[\protect\citeauthoryear{Coon et~al.}{2005}]{coon:05}
\begin{barticle}[auto:STB|2012/06/08|12:49:54]
\bauthor{\bsnm{Coon},~\bfnm{J.~J.}\binits{J.~J.}},
\bauthor{\bsnm{Syka},~\bfnm{J.~E.}\binits{J.~E.}},
\bauthor{\bsnm{Shabanowitz},~\bfnm{J.}\binits{J.}} \AND
\bauthor{\bsnm{Hunt},~\bfnm{D.}\binits{D.}}
(\byear{2005}).
\btitle{Tandem mass spectrometry for peptide and proteins sequence analysis}.
\bjournal{BioTechniques}
\bvolume{38}
\bpages{519--521}.
\bptok{imsref}%
\end{barticle}
\endbibitem

\bibitem[\protect\citeauthoryear{Dempster, Laird and Rubin}{1977}]{dempster:77}
\begin{barticle}[mr]
\bauthor{\bsnm{Dempster},~\bfnm{A.~P.}\binits{A.~P.}},
\bauthor{\bsnm{Laird},~\bfnm{N.~M.}\binits{N.~M.}} \AND
\bauthor{\bsnm{Rubin},~\bfnm{D.~B.}\binits{D.~B.}}
(\byear{1977}).
\btitle{Maximum likelihood from incomplete data via the {EM} algorithm}.
\bjournal{J. Roy. Statist. Soc. Ser. B}
\bvolume{39}
\bpages{1--38}.
\bid{issn={0035-9246}, mr={0501537}}
\bptnote{check related}%
\bptok{imsref}%
\end{barticle}
\endbibitem

\bibitem[\protect\citeauthoryear{Dongre et~al.}{1996}]{dongre:96}
\begin{barticle}[auto:STB|2012/06/08|12:49:54]
\bauthor{\bsnm{Dongre},~\bfnm{A.~R.}\binits{A.~R.}},
\bauthor{\bsnm{Johns},~\bfnm{J.~L.}\binits{J.~L.}},
\bauthor{\bsnm{Somogyi},~\bfnm{A.}\binits{A.}} \AND
\bauthor{\bsnm{Wysocki},~\bfnm{V.}\binits{V.}}
(\byear{1996}).
\btitle{Influence of peptide composition, gass-phase basicity, and chemical
modification on fragmentation efficiency: Evidence for the mobile proton
model}.
\bjournal{J. Am. Chem. Soc.}
\bvolume{118}
\bpages{8365--8374}.
\bptok{imsref}%
\end{barticle}
\endbibitem

\bibitem[\protect\citeauthoryear{Elias and Gygi}{2007}]{elias:07}
\begin{barticle}[auto:STB|2012/06/08|12:49:54]
\bauthor{\bsnm{Elias},~\bfnm{J.}\binits{J.}} \AND
\bauthor{\bsnm{Gygi},~\bfnm{S.}\binits{S.}}
(\byear{2007}).
\btitle{Target-decoy search strategy for increased confidence in large-scale
protein identifications by mass spectrometry}.
\bjournal{Nature Methods}
\bvolume{4}
\bpages{207--214}.
\bptok{imsref}%
\end{barticle}
\endbibitem

\bibitem[\protect\citeauthoryear{Eng, McCormack and Yates}{1994}]{eng:94}
\begin{barticle}[auto:STB|2012/06/08|12:49:54]
\bauthor{\bsnm{Eng},~\bfnm{J.}\binits{J.}},
\bauthor{\bsnm{McCormack},~\bfnm{A.}\binits{A.}} \AND
\bauthor{\bsnm{Yates},~\bfnm{J.~I.}\binits{J.~I.}}
(\byear{1994}).
\btitle{An approach to correlate tandem mass spectral data of peptides with
amino acid sequences in a protein database}.
\bjournal{J. Am. Soc. Mass Spectrom}
\bvolume{5}
\bpages{976--989}.
\bptok{imsref}%
\end{barticle}
\endbibitem

\bibitem[\protect\citeauthoryear{Fenyo and Beavis}{2003}]{fenyo:03}
\begin{barticle}[auto:STB|2012/06/08|12:49:54]
\bauthor{\bsnm{Fenyo},~\bfnm{D.}\binits{D.}} \AND
\bauthor{\bsnm{Beavis},~\bfnm{R.}\binits{R.}}
(\byear{2003}).
\btitle{A method for assessing the statistical significance
of mass spectrometry-based protein identifications using general scoring
schemes}.
\bjournal{Anal. Chem.}
\bvolume{75}
\bpages{768--774}.
\bptok{imsref}%
\end{barticle}
\endbibitem

\bibitem[\protect\citeauthoryear{Gerster et~al.}{2010}]{gerster:10}
\begin{barticle}[auto:STB|2012/06/08|12:49:54]
\bauthor{\bsnm{Gerster},~\bfnm{S.}\binits{S.}},
\bauthor{\bsnm{Qeli},~\bfnm{E.}\binits{E.}},
\bauthor{\bsnm{Ahrens},~\bfnm{C.~H.}\binits{C.~H.}} \AND
\bauthor{\bsnm{Buehlmann},~\bfnm{P.}\binits{P.}}
(\byear{2010}).
\btitle{Protein and gene model inference based on statistical
modeling in k-partite graphs}.
\bjournal{Proc. Natl. Acad. Sci. USA}
\bvolume{107}
\bpages{12101--12106}.
\bptok{imsref}%
\end{barticle}
\endbibitem

\bibitem[\protect\citeauthoryear{Hernandez, Muller and
Appel}{2006}]{hernandez:06}
\begin{barticle}[auto:STB|2012/06/08|12:49:54]
\bauthor{\bsnm{Hernandez},~\bfnm{P.}\binits{P.}},
\bauthor{\bsnm{Muller},~\bfnm{M.}\binits{M.}} \AND
\bauthor{\bsnm{Appel},~\bfnm{R.~D.}\binits{R.~D.}}
(\byear{2006}).
\btitle{Automated protein identification by tandem mass spectrometry: Issues
and strategies}.
\bjournal{Mass Spectrometry Reviews}
\bvolume{25}
\bpages{235--254}.
\bptok{imsref}%
\end{barticle}
\endbibitem

\bibitem[\protect\citeauthoryear{Keller et~al.}{2002a}]{keller:02b}
\begin{barticle}[pbm]
\bauthor{\bsnm{Keller},~\bfnm{Andrew}\binits{A.}},
\bauthor{\bsnm{Purvine},~\bfnm{Samuel}\binits{S.}},
\bauthor{\bsnm{Nesvizhskii},~\bfnm{Alexey~I.}\binits{A.~I.}},
\bauthor{\bsnm{Stolyar},~\bfnm{Sergey}\binits{S.}},
\bauthor{\bsnm{Goodlett},~\bfnm{David~R.}\binits{D.~R.}} \AND
\bauthor{\bsnm{Kolker},~\bfnm{Eugene}\binits{E.}}
(\byear{2002a}).
\btitle{Experimental protein mixture for validating tandem mass spectral
analysis}.
\bjournal{OMICS}
\bvolume{6}
\bpages{207--212}.
\bid{doi={10.1089/153623102760092805}, issn={1536-2310}, pmid={12143966}}
\bptok{imsref}%
\end{barticle}
\endbibitem

\bibitem[\protect\citeauthoryear{Keller et~al.}{2002b}]{keller:02}
\begin{barticle}[auto:STB|2012/06/08|12:49:54]
\bauthor{\bsnm{Keller},~\bfnm{A.}\binits{A.}},
\bauthor{\bsnm{Nesvizhskii},~\bfnm{A.}\binits{A.}},
\bauthor{\bsnm{Kolker},~\bfnm{E.}\binits{E.}} \AND
\bauthor{\bsnm{Aebersold},~\bfnm{R.}\binits{R.}}
(\byear{2002b}).
\btitle{Empirical statistical model to estimate the accuracy of peptide
identifications made by ms/ms and database search}.
\bjournal{Anal. Chem.}
\bvolume{74}
\bpages{5383--5392}.
\bptok{imsref}%
\end{barticle}
\endbibitem

\bibitem[\protect\citeauthoryear{Kinter and Sherman}{2000}]{kinter:00}
\begin{bbook}[auto:STB|2012/06/08|12:49:54]
\bauthor{\bsnm{Kinter},~\bfnm{M.}\binits{M.}} \AND
\bauthor{\bsnm{Sherman},~\bfnm{N.~E.}\binits{N.~E.}}
(\byear{2000}).
\btitle{Protein Sequencing and Identification Using Tandem Mass Spectrometry}.
\bpublisher{Wiley}, \baddress{New York}.
\bptok{imsref}%
\end{bbook}
\endbibitem

\bibitem[\protect\citeauthoryear{Klammer, Park and Noble}{2009}]{klammer:09}
\begin{barticle}[auto:STB|2012/06/08|12:49:54]
\bauthor{\bsnm{Klammer},~\bfnm{A.~A.}\binits{A.~A.}},
\bauthor{\bsnm{Park},~\bfnm{C.~Y.}\binits{C.~Y.}} \AND
\bauthor{\bsnm{Noble},~\bfnm{W.~S.}\binits{W.~S.}}
(\byear{2009}).
\btitle{Statistical calibration of the SEQUEST XCorr
function}.
\bjournal{Journal of Proteome Research}
\bvolume{8}
\bpages{2106--2113}.
\bptok{imsref}%
\end{barticle}
\endbibitem

\bibitem[\protect\citeauthoryear{Klammer et~al.}{2008}]{klammer:08}
\begin{barticle}[auto:STB|2012/06/08|12:49:54]
\bauthor{\bsnm{Klammer},~\bfnm{A.~A.}\binits{A.~A.}},
\bauthor{\bsnm{Reynolds},~\bfnm{S.}\binits{S.}},
\bauthor{\bsnm{MacCoss},~\bfnm{M.~J.}\binits{M.~J.}},
\bauthor{\bsnm{Bilmes},~\bfnm{J.}\binits{J.}} \AND
\bauthor{\bsnm{Noble},~\bfnm{W.}\binits{W.}}
(\byear{2008}).
\btitle{Modelling peptide fragmentation with dynamic Bayesian networks for
peptide identification}.
\bjournal{Bioinformatics}
\bvolume{24}
\bpages{i348--i356}.
\bptok{imsref}%
\end{barticle}
\endbibitem

\bibitem[\protect\citeauthoryear{Lam et~al.}{2007}]{lam:07}
\begin{barticle}[auto:STB|2012/06/08|12:49:54]
\bauthor{\bsnm{Lam},~\bfnm{H.}\binits{H.}},
\bauthor{\bsnm{Deutsch},~\bfnm{E.~W.}\binits{E.~W.}},
\bauthor{\bsnm{Eddes},~\bfnm{J.~S.}\binits{J.~S.}},
\bauthor{\bsnm{Eng},~\bfnm{J.~K.}\binits{J.~K.}},
\bauthor{\bsnm{King},~\bfnm{N.}\binits{N.}},
\bauthor{\bsnm{Stein},~\bfnm{S.~E.}\binits{S.~E.}} \AND
\bauthor{\bsnm{Aebersold},~\bfnm{R.}\binits{R.}}
(\byear{2007}).
\btitle{Development and validation of a spectral library
searching method for peptide identification from MS/MS}.
\bjournal{Proteomics}
\bvolume{7}
\bpages{655--667}.
\bptok{imsref}%
\end{barticle}
\endbibitem

\bibitem[\protect\citeauthoryear{Li, Eng and Stephens}{2012}]{li:12}
\begin{bmisc}[auto:STB|2012/06/08|12:49:54]
\bauthor{\bsnm{Li},~\bfnm{Q.}\binits{Q.}},
\bauthor{\bsnm{Eng},~\bfnm{J.~K.}\binits{J.~K.}} \AND
\bauthor{\bsnm{Stephens},~\bfnm{M.}\binits{M.}}
(\byear{2012}).
\bhowpublished{Supplement to ``A likelihood-based scoring method for peptide identification
using mass spectrometry.'' DOI:\doiurl{10.1214/12-AOAS568SUPP}.}
\bptok{imsref}%
\end{bmisc}
\endbibitem

\bibitem[\protect\citeauthoryear{Li, MacCoss and Stephens}{2010}]{li:10}
\begin{barticle}[mr]
\bauthor{\bsnm{Li},~\bfnm{Qunhua}\binits{Q.}},
\bauthor{\bsnm{MacCoss},~\bfnm{Michael~J.}\binits{M.~J.}} \AND
\bauthor{\bsnm{Stephens},~\bfnm{Matthew}\binits{M.}}
(\byear{2010}).
\btitle{A nested mixture model for protein identification using mass
spectrometry}.
\bjournal{Ann. Appl. Stat.}
\bvolume{4}
\bpages{962--987}.
\bid{doi={10.1214/09-AOAS316}, issn={1932-6157}, mr={2758429}}
\bptok{imsref}%
\end{barticle}
\endbibitem

\bibitem[\protect\citeauthoryear{Nesvizhskii and
Aebersold}{2004}]{nesvizhskii:04}
\begin{barticle}[auto:STB|2012/06/08|12:49:54]
\bauthor{\bsnm{Nesvizhskii},~\bfnm{A.~I.}\binits{A.~I.}} \AND
\bauthor{\bsnm{Aebersold},~\bfnm{R.}\binits{R.}}
(\byear{2004}).
\btitle{Analysis, statistical validation and dissermination of large-scale
proteomics datasets generated by tandem ms}.
\bjournal{Drug Discovery Today}
\bvolume{9}
\bpages{173--181}.
\bptok{imsref}%
\end{barticle}
\endbibitem

\bibitem[\protect\citeauthoryear{Nesvizhskii et~al.}{2003}]{nesvizhskii:03}
\begin{barticle}[auto:STB|2012/06/08|12:49:54]
\bauthor{\bsnm{Nesvizhskii},~\bfnm{A.}\binits{A.}},
\bauthor{\bsnm{Keller},~\bfnm{A.}\binits{A.}},
\bauthor{\bsnm{Kolker},~\bfnm{E.}\binits{E.}} \AND
\bauthor{\bsnm{Aebersold},~\bfnm{R.}\binits{R.}}
(\byear{2003}).
\btitle{A statistical model for identifying proteins by tandem mass
spectrometry}.
\bjournal{Anal. Chem.}
\bvolume{75}
\bpages{4646--4653}.
\bptok{imsref}%
\end{barticle}
\endbibitem

\bibitem[\protect\citeauthoryear{Sadygov, Liu and Yates}{2004}]{sadygov:04}
\begin{barticle}[auto:STB|2012/06/08|12:49:54]
\bauthor{\bsnm{Sadygov},~\bfnm{R.}\binits{R.}},
\bauthor{\bsnm{Liu},~\bfnm{H.}\binits{H.}} \AND
\bauthor{\bsnm{Yates},~\bfnm{J.}\binits{J.}}
(\byear{2004}).
\btitle{Statistical models for protein validation using tandem mass spectral
data and protein amino acid sequence databases}.
\bjournal{Anal. Chem.}
\bvolume{76}
\bpages{1664--1671}.
\bptok{imsref}%
\end{barticle}
\endbibitem

\bibitem[\protect\citeauthoryear{Shen et~al.}{2008}]{shen:08}
\begin{barticle}[auto:STB|2012/06/08|12:49:54]
\bauthor{\bsnm{Shen},~\bfnm{C.}\binits{C.}},
\bauthor{\bsnm{Wang},~\bfnm{Z.}\binits{Z.}},
\bauthor{\bsnm{Shankar},~\bfnm{G.}\binits{G.}},
\bauthor{\bsnm{Zhang},~\bfnm{X.}\binits{X.}} \AND
\bauthor{\bsnm{Li},~\bfnm{L.}\binits{L.}}
(\byear{2008}).
\btitle{A hierarchical statistical model to assess the confidence of peptides
and proteins inferred from tandem mass spectrometry}.
\bjournal{Bioinformatics}
\bvolume{24}
\bpages{202--208}.
\bptok{imsref}%
\end{barticle}
\endbibitem

\bibitem[\protect\citeauthoryear{Sun et~al.}{2007}]{sun:07}
\begin{barticle}[auto:STB|2012/06/08|12:49:54]
\bauthor{\bsnm{Sun},~\bfnm{S.}\binits{S.}},
\bauthor{\bsnm{Meyer-Arendt},~\bfnm{K.}\binits{K.}},
\bauthor{\bsnm{Eichelberger},~\bfnm{B.}\binits{B.}},
\bauthor{\bsnm{Brown},~\bfnm{R.}\binits{R.}},
\bauthor{\bsnm{Yen},~\bfnm{C.}\binits{C.}},
\bauthor{\bsnm{Old},~\bfnm{W.}\binits{W.}},
\bauthor{\bsnm{Pierce},~\bfnm{K.}\binits{K.}},
\bauthor{\bsnm{Cios},~\bfnm{K.}\binits{K.}},
\bauthor{\bsnm{Ahn},~\bfnm{N.~G.}\binits{N.~G.}} \AND
\bauthor{\bsnm{Resing},~\bfnm{K.~A.}\binits{K.~A.}}
(\byear{2007}).
\btitle{Improved validation of peptide ms/ms assignments using spectral
intensity prediction}.
\bjournal{Molecular and Cellular Proteomics}
\bvolume{6}
\bpages{1--17}.
\bptok{imsref}%
\end{barticle}
\endbibitem

\bibitem[\protect\citeauthoryear{Wan, Yang and Chen}{2006}]{wan:06}
\begin{barticle}[auto:STB|2012/06/08|12:49:54]
\bauthor{\bsnm{Wan},~\bfnm{Y.}\binits{Y.}},
\bauthor{\bsnm{Yang},~\bfnm{A.}\binits{A.}} \AND
\bauthor{\bsnm{Chen},~\bfnm{T.}\binits{T.}}
(\byear{2006}).
\btitle{PepHMM: A hidden Markov model based scoring function for mass
spectrometry database search}.
\bjournal{Anal. Chem.}
\bvolume{78}
\bpages{432--437}.
\bptok{imsref}%
\end{barticle}
\endbibitem

\bibitem[\protect\citeauthoryear{Wysocki et~al.}{2000}]{wysocki:00}
\begin{barticle}[auto:STB|2012/06/08|12:49:54]
\bauthor{\bsnm{Wysocki},~\bfnm{V.~H.}\binits{V.~H.}},
\bauthor{\bsnm{Tsaprsilis},~\bfnm{G.}\binits{G.}},
\bauthor{\bsnm{Smith},~\bfnm{L.}\binits{L.}} \AND
\bauthor{\bsnm{Breci},~\bfnm{L.~A.}\binits{L.~A.}}
(\byear{2000}).
\btitle{Mobile and localized protons: A framework for understanding peptide
dissociation}.
\bjournal{J. Mass Spectrom.}
\bvolume{35}
\bpages{1399--1406}.
\bptok{imsref}%
\end{barticle}
\endbibitem

\bibitem[\protect\citeauthoryear{Yen et~al.}{2011}]{yen:11}
\begin{barticle}[auto:STB|2012/06/08|12:49:54]
\bauthor{\bsnm{Yen},~\bfnm{C.}\binits{C.}},
\bauthor{\bsnm{Houel},~\bfnm{S.}\binits{S.}},
\bauthor{\bsnm{Ahn},~\bfnm{N.~G.}\binits{N.~G.}} \AND
\bauthor{\bsnm{Old},~\bfnm{W.}\binits{W.}}
(\byear{2011}).
\btitle{Spectrum-to-spectrum searching using a proteome-wide spectral library}.
\bjournal{Mol. Cell. Proteomics}
\bvolume{10}
\bpages{M111.007666}.
\bptok{imsref}%
\end{barticle}
\endbibitem

\bibitem[\protect\citeauthoryear{Yu et~al.}{2010}]{yu:10}
\begin{barticle}[auto:STB|2012/06/08|12:49:54]
\bauthor{\bsnm{Yu},~\bfnm{W.}\binits{W.}},
\bauthor{\bsnm{Taylor},~\bfnm{J.~A.}\binits{J.~A.}},
\bauthor{\bsnm{Davis},~\bfnm{M.~T.}\binits{M.~T.}},
\bauthor{\bsnm{Bonilla},~\bfnm{L.~E.}\binits{L.~E.}},
\bauthor{\bsnm{Lee},~\bfnm{K.~A.}\binits{K.~A.}},
\bauthor{\bsnm{Auger},~\bfnm{P.~L.}\binits{P.~L.}},
\bauthor{\bsnm{Farnsworth},~\bfnm{C.~C.}\binits{C.~C.}},
\bauthor{\bsnm{Welcher},~\bfnm{A.~A.}\binits{A.~A.}} \AND
\bauthor{\bsnm{Patterson},~\bfnm{S.~D.}\binits{S.~D.}}
(\byear{2010}).
\btitle{Maximizing the sensitivity and reliability of peptide
identification in large-scale proteomic experiments by harnessing multiple
search engines}.
\bjournal{Proteomics}
\bvolume{10}
\bpages{1172--1189}.
\bptok{imsref}%
\end{barticle}
\endbibitem

\bibitem[\protect\citeauthoryear{Zhang}{2004}]{zhang:04}
\begin{barticle}[pbm]
\bauthor{\bsnm{Zhang},~\bfnm{Zhongqi}\binits{Z.}}
(\byear{2004}).
\btitle{Prediction of low-energy collision-induced dissociation spectra of
peptides}.
\bjournal{Anal. Chem.}
\bvolume{76}
\bpages{3908--3922}.
\bid{doi={10.1021/ac049951b}, issn={0003-2700}, pmid={15253624}}
\bptok{imsref}%
\end{barticle}
\endbibitem

\bibitem[\protect\citeauthoryear{Zhang}{2005}]{zhang:05}
\begin{barticle}[pbm]
\bauthor{\bsnm{Zhang},~\bfnm{Zhongqi}\binits{Z.}}
(\byear{2005}).
\btitle{Prediction of low-energy collision-induced dissociation spectra of
peptides with three or more charges}.
\bjournal{Anal. Chem.}
\bvolume{77}
\bpages{6364--6373}.
\bid{doi={10.1021/ac050857k}, issn={0003-2700}, pmid={16194101}}
\bptok{imsref}%
\end{barticle}
\endbibitem

\end{thebibliography}
\end{document}